\documentclass[a4paper]{article}

\usepackage{cite}
\usepackage{INTERSPEECH2019}
\usepackage{amsmath}
\usepackage{amssymb}
\usepackage{math}
\usepackage{comment}
\usepackage{graphicx}
\usepackage{caption}
\usepackage{soul}
\usepackage[bottom]{footmisc}
\usepackage{float}
\usepackage[utf8]{inputenc}
\usepackage{textpos}
\newcommand{\minus}{\scalebox{0.75}[1.0]{$-$}}
\newcommand{\Test}[1]{\expandafter\hat#1}
\def\Thetamat{\mbox{\boldmath $\Theta$}}
\def\Wmat{\mbox{\boldmath $W$}}
\def\Hmat{\mbox{\boldmath $H$}}
\def\Xmat{\mbox{\boldmath $X$}}

\title{A Statistically Principled and Computationally Efficient Approach to Speech Enhancement using Variational Autoencoders}
\name{Manuel Pariente \thanks{Experiments presented in this paper were carried out using the Grid’5000 testbed, supported by a scientific interest group hosted by Inria and including CNRS, RENATER and several Universities as well as other organizations (see \texttt{https://www.grid5000.fr}).} \qquad Antoine Deleforge \qquad Emmanuel Vincent}
\address{
  Universitee de Lorraine, CNRS, Inria, LORIA, F-54000 Nancy, France}
\email{manuel.pariente@univ-lorraine.fr, \{antoine.deleforge, emmanuel.vincent\}@inria.fr}

\begin{document}

\maketitle
\begin{textblock*}{3cm}(11.1cm,-6.2cm)
   \mbox{\textit{SUBMITTED TO INTERSPEECH 2019}}
\end{textblock*}

\begin{abstract}
Recent studies have explored the use of deep generative models of speech spectra based of variational autoencoders (VAEs), combined with unsupervised noise models, to perform speech enhancement. These studies developed iterative algorithms involving either Gibbs sampling or gradient descent at each step, making them computationally expensive. This paper proposes a variational inference method to iteratively estimate the power spectrogram of the clean speech. Our main contribution is the analytical derivation of the variational steps in which the encoder of the pre-learned VAE can be used to estimate the variational approximation of the true posterior distribution, using the very same assumption made to train VAEs. Experiments show that the proposed method produces results on par with the aforementioned iterative methods using sampling, while decreasing the computational cost by a factor 36 to reach a given performance.
\end{abstract}

\section{Introduction}
Speech enhancement is the problem of extracting a speech source from a noisy recording \cite{Loizou2013SE, VincentBook2018}. It is commonly tackled by discriminative methods using deep neural networks (DNNs) to predict time-frequency masks \cite{Weninger2014, Hershey_2016DC}, clean power spectrograms\cite{Lu2013SpeechEB} or to model the signal variance \cite{Nugraha2016a} from spectrograms of mixtures. While the performances of these methods are satisfactory, their generalization capabilities are limited due to their supervised nature. Indeed, DNNs trained this way are specific to a task and a recording condition. Instead, generative approaches do not suffer from this drawback. Once a source model has been learned, the statistical model can account for the different tasks and recording conditions.

Non-negative matrix factorization (NMF) \cite{Ozerov2010} is a popular method for speech enhancement in which power or amplitude spectrograms are approximated as the product of two non-negative matrices. While certain forms of NMF can be interpreted as generative, they suffer from their linearity assumptions.
Very recent studies have combined the modelling power of DNN with statistical approaches in order to design unsupervised speech enhancement and source separation methods in both single-channel \cite{Leglaive2018Single,Leglaive2019Alpha, Bando2018} and multichannel scenarios \cite{Leglaive2018Multi, Sekiguchi2018, Kameoka2018, Seki2018, Li2018}.
The main idea of these studies is to use variational autoencoders~(VAEs) \cite{Kingma2014} to replace the NMF generative model, thus benefiting from DNNs' representational power. Combined with an observation model, Expectation-Maximization (EM) \cite{MCEM1990} or Bayesian inference \cite{Bishop2006} algorithms can be derived to iteratively estimate individual source spectra from a given mixture.

In \cite{Leglaive2018Single, Leglaive2019Alpha, Bando2018, Leglaive2018Multi, Sekiguchi2018}, speech enhancement is performed using a pretrained VAE-based generative model of speech spectra combined with an unsupervised NMF model for the noise. Inference of the clean speech involves using the Metropolis-Hastings algorithm \cite{MC2005} to estimate an intractable distribution over the VAE's latent space.
Similar approaches applied to source separation are developed in \cite{Kameoka2018, Seki2018}, where the latent variables are updated using backpropagation.
While the algorithms analytically derived in \cite{Leglaive2018Single,Leglaive2019Alpha, Bando2018, Leglaive2018Multi, Sekiguchi2018, Kameoka2018, Seki2018, Li2018} show promising performances, sampling and backpropagation methods are computationally expensive. 

In order to train a VAE, an encoder is jointly learned with the generative model. The role of the encoder is to approximate the true, but intractable, posterior. Once the VAE is trained, the encoder could in principle be used to approximate the true posterior. Interestingly, \cite{Li2018} proposes a heuristic algorithm for multichannel source separation based on \cite{Kameoka2018} which uses this property of the encoder to achieve computational efficiency. However this inference algorithm is not statistically principled.

In this study, we present a Bayesian single-channel speech enhancement algorithm in which a VAE is used as generative model of speech spectra, and the noise is modelled using NMF. After unsupervised training of the speech model, the inference constits in iteratively updating the source estimates using Wiener filtering and updating the corresponding latent representation using the encoder. No sampling or backpropagation being required, our method is significantly faster than in
 \cite{Leglaive2018Single,Leglaive2019Alpha, Bando2018, Leglaive2018Multi, Sekiguchi2018, Kameoka2018, Seki2018}.
The use of the encoder differs from \cite{Li2018}, and, unlike it, it is not heuristic but motivated by the same variational approximation as the one used to train VAEs.
To the best of our knowledge, this is the first method to propose a statistically principled approach re-using the probabilistic encoder as a posterior approximator.

We first introduce the VAE framework and its application to spectrogram modelling. We then detail our assumptions and our inference algorithm. Experiments and results will be presented before we finally conclude.

\section{Model} \label{sec:Model}
Given an observation of clean speech embedded in noise, the goal of speech enhancement is to produce an estimate of the clean speech. Our approach is to use a VAE as the generative model of clean speech spectra to infer this estimate. We will first describe the VAE framework and its training procedure in Section \ref{ssec:VAE-model} and \ref{ssec:VAE-training} before introducing the observation model in Section \ref{ssec:Mixt-model}.
\subsection{Generative speech model} \label{ssec:VAE-model}
A common approach to modeling sound sources is to assume the short time Fourier Transform (STFT) coefficients are proper complex Gaussian random variables. With $f$ denoting the frequency index and $t$ the time-frame index, for all $(f, t) \in \{0,... ,F-1\}\times \{0,... , N-1\}$, each source STFT coefficient $s_{ft} \in \mathbb{C}$ independently follows
\begin{equation}
    \label{eq:source model}
    s_{ft} \sim \mathcal{N}_c(0,\,\sigma^{2}_{ft}).
\end{equation}
Several frameworks have been used to model the variance $\sigma^{2}_{ft}$ of this distribution \cite{Vincent2010, VincentBook2018}. To this end, we choose to use the VAE framework, which uses the representational power of DNNs to build generative models. It consists in mapping a zero-mean unit Gaussian random variable $\zvect_t \in \mathbb{R}^L$ through a DNN, into the desired output distribution. In our setting, we use it to represent $\sigma^{2}_{ft}$. The generative model can be written as
\begin{align}
    \label{eq:p(z)}
    \zvect_t &\sim \mathcal{N}(\zerovect, \Imat), \\
    \label{eq:p(s|z)}
    s_{ft} | \zvect_{t}; \theta &\sim \mathcal{N}_c(0, \sigma_{f}^2(\zvect_{t})),
\end{align}
where $\sigma_{f}^2 : \mathbb{R}^L \mapsto \mathbb{R}_{+}$ is a non-linear function implemented by a DNN with parameters \thetavect. Each $\sigma_{f}^2$ is one neuron of a F-dimensional output layer of a single DNN.
\subsection{VAE training and posterior approximation} \label{ssec:VAE-training}
The VAE is trained on clean speech data by maximizing the likelihood $p_{\theta}(\svect) = \prod_t p_{\theta}(\svect_t)$, with $\svect_t = \left[s_{1t},..., s_{Ft}\right]^T$. To do so, we introduce a variational approximation of $p_{\theta}(\zvect_t|\svect_t)$ , $q_{\phi}(\zvect_t|\svect_t)$, parametrized by $\phivect$. We have
\begin{align}
    \label{eq:LL_VAE}
    \log p_{\theta}(\svect_t) &= \mathcal{D}_{KL}(q_{\phi}(\zvect_t | \svect_t) || p_{\theta}(\zvect_t | \svect_t)) + \mathcal{L}(\thetavect, \phivect; \svect_t) \\
    \label{eq:ELBO_VAE}
    \mathcal{L}(\thetavect, \phivect; \svect_t) &=\mathbb{E}_{q_{\phi}}[\log p_{\theta}(\svect_t|\zvect_t)] - \mathcal{D}_{KL}(q_{\phi}(\zvect_t | \svect_t) || p(\zvect_t)),
\end{align}
where $\mathcal{D}_{KL}(q||p) = - \mathbb{E}_{q}[\log (p/q)]$ denotes the Kullback-Leibler (KL) divergence. The KL term on the left hand side of \eqref{eq:LL_VAE} being always positive, $\mathcal{L}(\thetavect, \phivect; \svect_t)$ is a lower bound of the marginal log-likelihood $\log p_{\theta}(\svect_t)$.
We can thus maximize the individual $\mathcal{L}(\thetavect, \phivect; \svect_t)$ with respect to $\thetavect$ and $\phivect$ in order to maximize the log-likelihood $\log p_{\theta}(\svect)$.
For all $(l, t) \in \{0,... ,L-1\}\times \{0,... , N-1\}$, $q_{\phi}(z_{lt}|\svect_t)$ is defined as in \cite{Leglaive2018Single}
\begin{equation}
    \label{eq:approximate posterior}
    z_{l, t} | \svect_t;\phi  \sim \mathcal{N}(\tilde{\mu}_{l}({|\svect_t}|^2), \tilde{\sigma}_{l}^2({|\svect_t|^2})), 
\end{equation}
where $\tilde{\mu}_{l}$ and $\tilde{\sigma}_{l}^2$ are DNN-based functions parametrized by $\phivect$.
Finally, we can use \eqref{eq:p(z)}, \eqref{eq:p(s|z)} and \eqref{eq:approximate posterior} to write \eqref{eq:ELBO_VAE} as 
\begin{align}
    \label{eq:loss function}
    \mathcal{L}(\thetavect, \phivect; \svect) &\overset{c}{=} \sum_{f} \mathbb{E}_{q_{\phi}(\zvect_t|s_{ft})}d_{IS}(|s_{ft}|^2, \sigma_f^2(\zvect_{t})) \\
    & + \dfrac{1}{2}\sum_{l} \log({\tilde{\sigma}^2}_{l}(|\svect_t|^2)) - \tilde{\mu}^2_{l}(|\svect_t|^2) - \tilde{\sigma}^2_{l}(|\svect_t|^2),\nonumber 
\end{align}
where $d_{IS}(x,y) = x/y - \log(x/y) - 1$ denotes the Itakura-Saito divergence, $\zvect_t = \left[z_{1t},..., z_{Lt}\right]^T$ and $\overset{c}{=}$ denotes equality up to a constant. Using the so-called reparametrization trick \cite{Kingma2014}, we can maximize $\mathcal{L}(\thetavect, \phivect; \svect)$ using gradient descent optimization algorithms \cite{ruder2016GradDesc}.

The variational distribution $q_{\phi}(\zvect_t|\svect_t)$ is usually considered as an unwanted by-product of the training procedure of the VAE, only introduced to learn $p_{\theta}(\svect_t|\zvect_t)$.
However, by examining carefully \eqref{eq:LL_VAE}, we can see that while maximizing $\mathcal{L}(\thetavect, \phivect; \svect_t)$ can only increase the log-likelihood, it also decreases the KL divergence between the true posterior and its variational approximation.
In other words, in addition to the generative speech model, the VAE provides a variational approximation of the true posterior. This feature  will play a key role in the inference algorithm proposed in Section \ref{sec:Inference}.

\subsection{Noisy speech model} \label{ssec:Mixt-model}
As in \cite{Leglaive2018Single}, the STFT coefficients $n_{ft}$ of the noise are modelled with a rank-$K$ NMF Gaussian model. For all $(f,t)$ 
\begin{equation}
    \label{eq:NMF-noise}
    n_{ft} \sim \mathcal{N}_c(0, (\Wmat \Hmat)_{ft}),
\end{equation}
with $\Wmat \in \mathbb{R}_+^{F \times K}$, $\Hmat \in \mathbb{R}_+^{K \times N}$.

Finally, independently for all $(f,t)$, the single-channel observation of the noisy speech is modeled by
\begin{equation}
    \label{eq:mixture-model}
    x_{ft} = s_{ft} + n_{ft},
\end{equation}
where $s_{ft}$ and $n_{ft}$ are independent.

\section{Inference} \label{sec:Inference}
Given a noisy speech signal $\Xmat = \{x_{ft}\}_{(ft)}$, our goal is now to maximize the likelihood of $\Xmat$ given the mixture model \eqref{eq:mixture-model}, the generative model of speech \eqref{eq:p(z)}, \eqref{eq:p(s|z)} and the NMF model \eqref{eq:NMF-noise}.
With $\nvect_t = \left[n_{1t},..., n_{Ft}\right]^T$ and $\Hvect_t = \left[H_{1t},..., H_{Kt}\right]^T$, we consider $\yvect_t = \{\svect_t, \nvect_t, \zvect_t\}$ to be the set of latent variables and $\Thetamat_t = \{\Wmat, \Hmat_t\}$ the parameters of the model. We introduce $r(\yvect_t)$, a variational approximation of $p(\yvect_t | \xvect_t;\Thetamat_t)$. We have
\begin{align}
    \nonumber
    \log p(\xvect_t; \Thetamat_t) &= \mathcal{D}_{KL}(r(\yvect_t) || p(\yvect_t | \xvect_t; \Thetamat_t)) + \mathcal{L}(r, \Thetamat_t)\\
    \mathcal{L}(r, \Thetamat_t) &= \mathbb{E}_{r(\yvect_t; \Thetamat)}\left[\log \frac {p(\xvect_t,\yvect_t; \Thetamat_t)}{r(\yvect_t)}\right].
\end{align}
We suppose that the variational distribution $r(\yvect_t)$ factorizes as 
\begin{equation}
   \label{eq:mean field} 
   r(\svect_t, \nvect_t, \zvect_t) = r(\svect_t, \nvect_t)r(\zvect_t) = \prod_f r(s_{ft}, n_{ft}) \prod_l r(z_{lt}).
\end{equation}
We can then iteratively maximize $\mathcal{L}(r, \Thetamat_t)$ with respect to the factorized distributions and the NMF parameters. The variational distributions' updates are given by (10.9) in \cite{Bishop2006} 
\begin{align}
    \label{eq:sb-step}
    \log r(s_{ft}, n_{ft}) &\overset{c}{=} \mathbb{E}_{r(\zvect_t)}\log p(x_{ft}, s_{ft}, n_{ft}, \zvect_t ; \Thetamat_{ft}) \\
    \label{eq:z-step}
    \log r(z_{lt}) &\overset{c}{=} \mathbb{E}_{r(\svect_t, \nvect_t)}\log p(\xvect_t, \svect_t, \nvect_t, z_{lt} ; \Thetamat_t).
\end{align}
$\Wmat$ and $\Hmat$ can be updated by maximizing the following:
\begin{equation}
	\label{eq:M-cost}
    \mathcal{Q}(\Thetamat, \Thetamat^{old}) \overset{c}{=} \mathbb{E}_{r(\yvect; \Thetamat^{old})}[\log p(\xvect, \yvect; \Thetamat)].
\end{equation}
We detail those updates below, and in the supporting document \cite{SupDoc}.

\subsection{E-(s,n) step}
We define $\sigma^2_{n, ft} = (\Wmat\Hmat)_{ft}$ to make the notation less cluttered. Using \eqref{eq:sb-step}, we find (see \cite{SupDoc})
\begin{equation}
    \label{eq:r(s,b)}
    r(\svect_{t}, \nvect_{t}) \sim \prod_f \mathcal{N}_c(\muvect_{ft}, \Sigmamat_{ft}) = \mathcal{N}_c(\muvect_{t}, \Sigmamat_{t}), 
\end{equation}
where $\muvect_{ft}$ and $\Sigmamat_{ft}$ are defined as
\begin{gather}
    \label{eq:var_musig}
    \Sigmamat_{ft} = \frac{\gamma^2_{ft}\sigma^2_{n, ft}}{\gamma^2_{ft} + \sigma^2_{n, ft}}
    \begin{bmatrix} 1 &\minus 1 \\ \minus 1 & 1 \end{bmatrix},
    \quad
    \muvect_{ft} = \frac{x_{ft}}{\gamma^2_{ft} + \sigma^2_{n, ft}}  \begin{bmatrix} \gamma^2_{ft} \\ \sigma^2_{n, ft} \end{bmatrix},
\end{gather}
with
\begin{equation}
    \label{eq:gamma}
    \frac{1}{\gamma^2_{ft}} = \mathbb{E}_{r(\zvect_t)}\Big[1/{\sigma_f^2(\zvect_t)}\Big] \approx \sum_{d=1}^{D} \Big[1/{\sigma_f^2(\zvect_t^{(d)})}\Big],
\end{equation}
where $\{\zvect_n^{(d)}\}_{d=1,..,D}$ are randomly drawn from $r(\zvect_t)$. 

We note $[\muvect_{s, t}, \muvect_{n, t}] = \muvect_{t}$ and define $\Sigmamat_{ss, t}$ and $\Sigmamat_{nn, t}$ to be the diagonal terms of $\Sigmamat_{t}$.

\subsection{E-z step}
We can compute $r(\zvect_t)$ using \eqref{eq:z-step} (see \cite{SupDoc}), we get
\begin{align}
	\nonumber
    \log r(\zvect_t) &\overset{c}{=} \log p(\zvect_t) + \log p(\svect_t= \left(|\muvect_{s, t}|^2 + \Sigmamat_{ss, t}\right)^{\frac{1}{2}} | \zvect_t) \\
    \label{eq:r(z) inverted}
    &= \log p(\zvect_t | \svect_t =\left(|\muvect_{s, t}|^2 + \Sigmamat_{ss, t}\right)^{\frac{1}{2}}),
\end{align}
where \eqref{eq:r(z) inverted} was obtained using Bayes' theorem.
The VAE assumes that $p_{\theta}(\zvect_t | \svect_t)$ can be appproximated by $q_{\phi}(\zvect_t| \svect_t)$. We further assume that this still holds for $\svect_t$ of the form $\left(|\muvect_{s, t}|^2 + \Sigmamat_{ss, t}\right)^{\frac{1}{2}}$. That is to say
\begin{align}
    \label{eq:posterior approximation}
    r(z_{lt}) &\approx q_{\phi}(z_{lt}|\svect_t =\left(|\muvect_{s, t}|^2 + \Sigmamat_{ss, t}\right)^{\frac{1}{2}}) \\
    \label{eq:variational z-update}
     &= \mathcal{N}(\tilde{\mu}_{l}(|\muvect_{s, t}|^2 + \Sigmamat_{ss, t}), \tilde{\sigma}_{l}^2(|\muvect_{s, t}|^2 + \Sigmamat_{ss, t})).
\end{align}

\subsection{M-step}
We now maximize $\mathcal{L}(r, \Thetamat_t)$ with respect to $\Thetamat_t$ using \eqref{eq:M-cost}.
With $\Vvect \in \mathbb{R}_+^{F \times N}$ defined as  $(\Vvect)_{t} = |\muvect_{n, t}|^2 + \Sigmamat_{nn, t}$,  and $\odot$ denoting element-wise matrix multiplication and exponentiation, we obtain multiplicative updates as in \cite{Fevotte2009}:
\begin{equation}
    \label{eq:H-up}
    \Hmat \gets \Hmat \odot \frac{\Wmat^T \Big( (\Wmat\Hmat)^{\odot -2} \odot \Vmat \Big)}{\Wmat^T(\Wmat\Hmat)^{\odot-1}},
\end{equation}
\begin{equation}
    \label{eq:W-up}
    \Wmat \gets \Wmat \odot \frac{\Big( (\Wmat\Hmat)^{\odot -2} \odot \Vmat \Big)\Hmat^T}{(\Wmat\Hmat)^{\odot-1}\Hmat^T}.
\end{equation}

\subsection{Speech Reconstruction} \label{ssec: Speech recon}
Let $\Thetamat^{\star} = \{\Wmat^{\star}, \Hmat^{\star}\}$ and $r^{\star}(\svect, \bvect, \zvect) = r^{\star}(\svect, \bvect)r^{\star}(\zvect)$ be respectively the set of NMF parameters and the variational distribution estimated by the proposed algorithm. The final estimate of the source is given, as in \cite{Leglaive2018Single}, by 
\begin{align}
    \label{eq:posterior_mean}
    \Test{s_{ft}} &= \mathbb{E}_{p(s_{ft}|x_{ft};\Thetamat^{\star})}[s_{ft}] \\
    \label{eq:exp_wiener}
    &= \mathbb{E}_{p(\zvect_{t}|x_{ft};\Thetamat^{\star})}\Bigg[\frac{\sigma^2_f(\zvect_t)}{\sigma^2_f(\zvect_t) + (\Wmat\Hmat)_{ft}}\Bigg]x_{ft}.
\end{align}
There are three ways to evaluate this estimate.
The straightforward way is to replace $p(s_{ft}|x_{ft};\Thetamat^{\star})$  in \eqref{eq:posterior_mean} by its variational approximation $r^{\star}(s_{ft})$. The expectation over $p(\zvect_{t}|x_{ft};\Thetamat^{\star})$ can also be approximated using the Metropolis Hastings (MH) algorithm as in \cite{Leglaive2018Single}, using the mean of $r^{\star}(\zvect_t)$ as the initial sample. And finally we can compute the expectation in \eqref{eq:exp_wiener} by replacing $p(\zvect_{t}|x_{ft};\Thetamat^{\star})$ by its variational approximation $r^{\star}(\zvect_t)$. We will respectively refer to these methods as \textit{S-Wiener},  \textit{MH-Wiener} and  \textit{Z-Wiener} for ease of referencing.\\
\textit{Metropolis-Hastings :} At the $m$-th iteration of the Metropolis-Hastings algorithm, we draw a random sample for each $t$ according to 
\begin{equation}
    \label{eq:Metropolis Hastings}
    \zvect_t | \zvect^{(m-1)}_t; \epsilon_{mh}^2 \sim \mathcal{N}(\zvect^{(m-1)}_t, \epsilon_{mh}^2\Imat).
\end{equation}
The acceptance probability can be computed as 
\begin{equation}
    \label{eq:acceptance}
    \alpha = \min \Bigg(1, \frac{p(\xvect_t | \zvect_t; \Thetamat^{\star})p(\zvect_t)}{p(\xvect_t | \zvect^{(m-1)}_t; \Thetamat^{\star})p(\zvect^{(m-1)}_t)}\Bigg).
\end{equation}
We accept the new sample and set $\zvect^{(m)}_t = \zvect_t$ only if $u$, drawn from a uniform distribution $\mathcal{U}([0, 1])$, is smaller than $\alpha$. We keep only the last $R$ samples to compute $\hat{s}_{ft}$ from \eqref{eq:exp_wiener} and discard the samples drawn during the burn-in period.   

\section{Experimental evaluation}
\subsection{Experimental settings} \label{ssec:Exp setting}
\textit{Dataset}:
As in \cite{Leglaive2018Single, Leglaive2019Alpha}, we use the the TIMIT corpus \cite{TIMIT} to train the clean speech model. At inference time, we mix speech signals from the TIMIT test set with noise signals from the DEMAND corpus \cite{DEMAND} at 0dB signal-to-noise ratio, according to the file provided in in the original implementation of \cite{Leglaive2018Single} \footnote{https://gitlab.inria.fr/sileglai/mlsp\-2018}. Note that both speakers and utterances are different from those in the training set. 

\textit{VAE's architecture and training}:
The architecture of the VAE is the same as in \cite{Leglaive2018Single}. The hyperbolic tangent of an input 128-dimensional linear layer passes through two L-dimensional linear layers to estimate $\tilde{\mu}_{l}({|\svect_t}|^2)$ and $\log \tilde{\sigma}_{l}^2({|\svect_t|^2})$, from which $\zvect_t$ is sampled using the reparametrization trick. $\zvect_t$ is then fed to a 128-dimensional linear layer with hyperbolic tangent activation followed by an linear output layer predicting  $\log \sigma_{f}^2(\zvect_{t})$.  
For training, we computed the STFT of the utterances with a 1024-points sine window and a hop size of 256 points. 20$\%$ of the training set is held for validation. A VAE with $L=64$ is trained with the Adam optimizer \cite{Adam}, using a learning rate of $10^{-3}$ and a batch size of 128. Training stops if no improvement is seen on the validation loss for 10 epochs. 

\textit{Baseline methods}:
We compare our method to two baselines.
The first baseline, developed in \cite{Leglaive2018Single}, shares the same statistical assumptions made in Section \ref{sec:Model} except for the inclusion of a gain factor in \eqref{eq:mixture-model}. The inference is performed using a Monte Carlo Expectation-Maximization (MCEM) algorithm in which MH sampling is used to estimate the true posterior. We will refer to this method as \textit{MCEM}.
The second baseline is a heuristic we introduce to evaluate the impact of the covariance term $\Sigmamat_{ss, t}$ in \eqref{eq:variational z-update} on the convergence of the algorithm. We simply remove $\Sigmamat_{ss, t}$ from \eqref{eq:variational z-update} and the rest of the algorithm remains unchanged. Note that this is similar to the heuristic developped in \cite{Li2018} for determined multichannel source separation, only adapted to single-channel speech enhancement.
\setlength{\textfloatsep}{10pt plus 1.0pt minus 2.0pt}
\begin{figure}[t]
  \centering
  \includegraphics[scale=0.52]{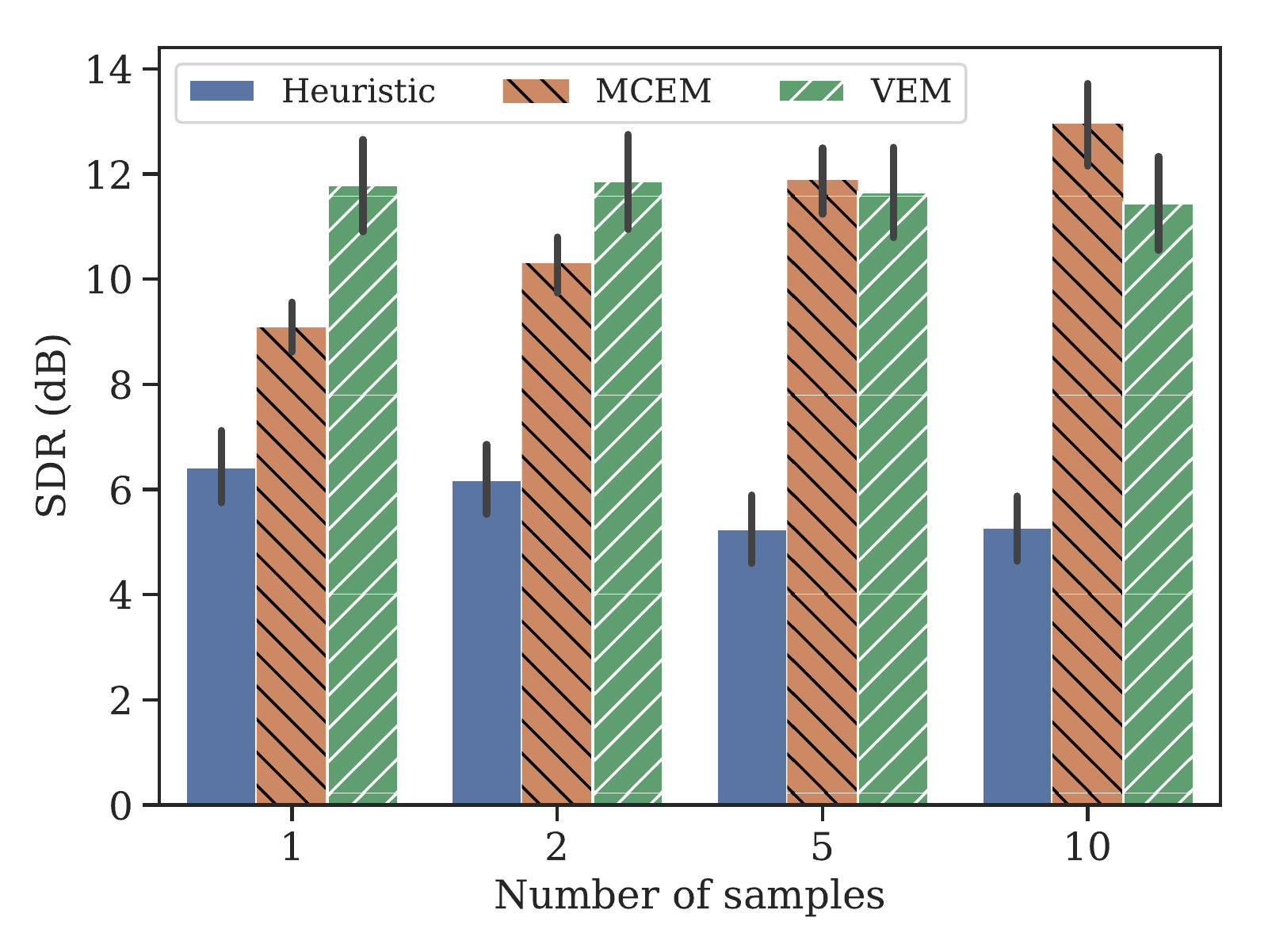}
  \caption{SDR as a function of number of samples M for an input SNR of 0dB. using MH for reconstruction. Heuristic, MCEM and VEM methods are defined in Section \ref{ssec:Exp setting}. Errors bars indicate 95\% confidence interval.}
  \label{fig:sdr_plot}
\end{figure} 

\textit{Inference Setting}:
For fair comparaison, we use the same settings for the three methods : the rank of the NMF model is set to $K=10$, the latent dimension of the VAE is set to $L=64$ and the same stopping criterion is used as in \cite{Leglaive2018Single}. $\zvect_t$ is initialized based on $q_{\phi}(\zvect_t|\xvect_t)$ and for the final speech estimate, equation \eqref{eq:exp_wiener} is approximated using the last 25 samples of a 100-iteration MH sampling with $\epsilon^2=0.01$.
In \cite{Leglaive2018Single}, for $R$ samples used, the total number of draws is $4R$. The methods will be compared for a same number of samples actually used to estimate the expected values : if $D$ samples are used to evaluate \eqref{eq:gamma},  $4D$ samples will be drawn in the MCEM case.
\subsection{Results}
\textit{Experiment 1}:
We first compare the performances of the three methods on the test set described above, in terms of Signal to Distorsion Ration (SDR) \cite{Evincent2006BSS_eval}, computed using the \textit{mir\_eval}\footnote{https://github.com/craffel/mir\_eval} toolbox. The comparison is done for different numbers of samples $D \in \{1, 2, 5, 10\}$.
As can be seen in Fig. \ref{fig:sdr_plot}, the heuristic algorithm consistently performs worse than both other methods, proving the importance of the covariance term $\Sigmamat_{ss, t}$ in \eqref{eq:variational z-update}. We also see that the performances of the proposed algorithm do not depend on the number of samples drawn to approximate \eqref{eq:gamma}, contrary to the MCEM algorithm which presents poor performances for $R=1$ and outperforms the proposed method for $R=10$. The superiority of the MCEM algorithm for a large number of samples is not surprising due to the approximation involved in the VEM approach.

\begin{figure}[t]
  \centering
  \includegraphics[scale=0.52]{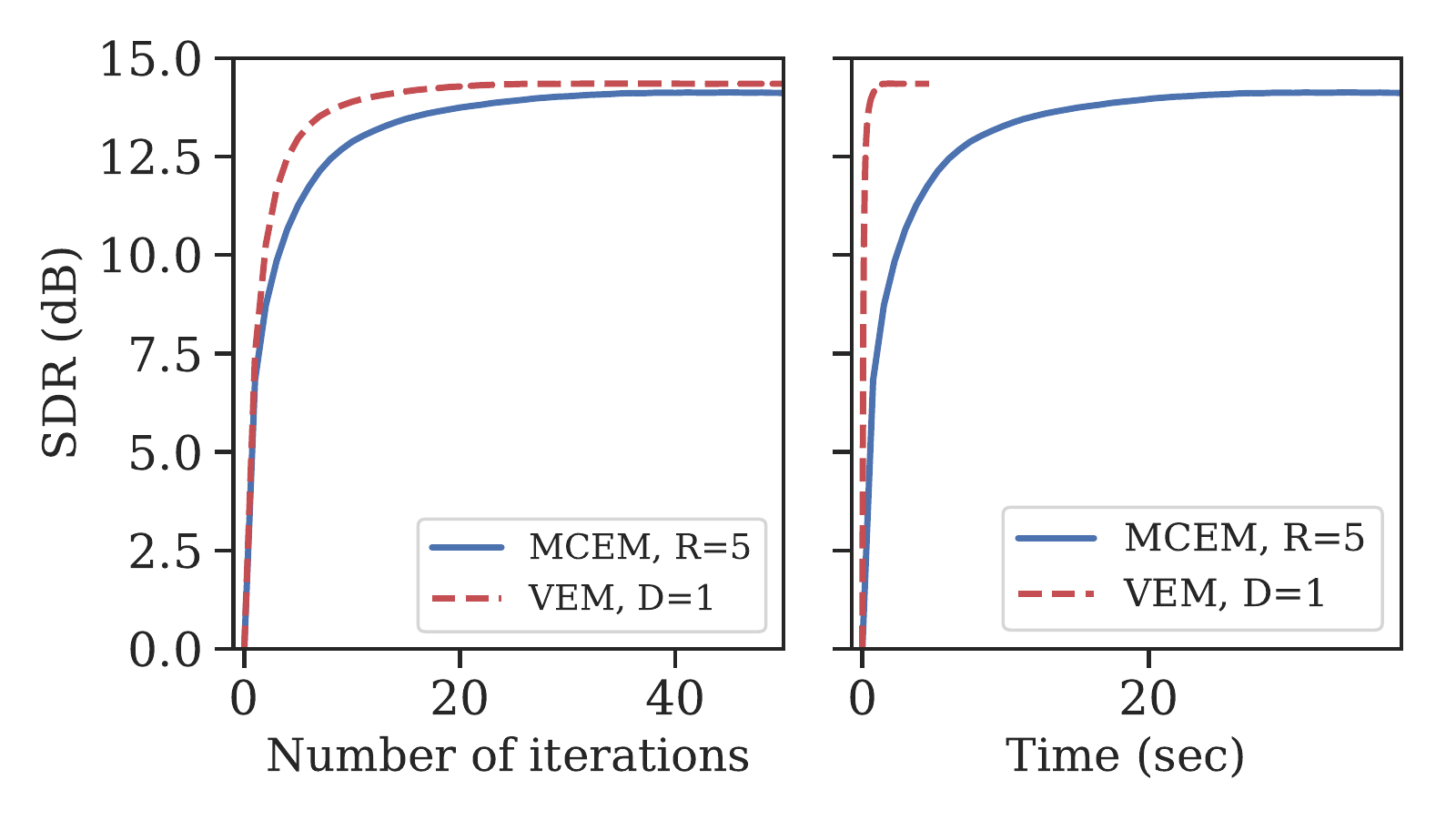}
  \caption{Average convergence speed. SDR as a function of (left) number of iteration and (right) absolute time.}
  \label{fig:sdr_time}
\end{figure}

\textit{Experiment 2}:
We then compare the SDR achieved by the MCEM and VEM methods as a function of the computational time, in terms of number of iteration and absolute time. The number of samples $D$ was set to 1 for the VEM, and $R$ to 5 for the MCEM so as to achieve a similar SDR at convergence (see Fig. \ref{fig:sdr_plot}). 
At each iteration of both methods, we estimate the speech spectra by approximating \eqref{eq:exp_wiener} with MH sampling and we compute the SDR. This is done for each test utterance. Each individual SDR curve is then padded with the SDR obtained at the last iteration and all SDR curves are averaged to produce the left graph in Fig. \ref{fig:sdr_time}. We observe that the proposed method converges faster which suggests that using the encoder allows for bigger jumps in the latent space than the sampling method with a small number of samples. On a 4-core i7-8650U CPU, one iteration takes on average 55 ms and 753 ms for the VEM and the MCEM algorithm respectively. Using these numbers, we can convert the SDR as a function of the number of iterations to the SDR as a function of time. This is shown in the right part of Fig. \ref{fig:sdr_time} where the superiority of the proposed algorithmm in terms of absolute speed is clear. To compare the execution times, for each utterance and for both algorithms we compute the number of iteration needed to reach the final SDR up to 0.5dB tolerance. The ratio between the number of iterations multiplied by the time per iteration ratio yields an average computational cost decrease factor of 36 to reach the same performance.
\setlength{\textfloatsep}{10pt plus 1.0pt minus 2.0pt}

\textit{Experiment 3}:
In this experiment, we compare the three different ways of computing the source STFT coefficients in \eqref{eq:posterior_mean}. We present the mean SDR after convergence for the three possible methods in Table \ref{table:recon_table}.
We can see that MH sampling consistently outperforms the other methods, suggesting that a better posterior approximation benefits to the quality of the reconstruction.

\begin{table}[H]
	\centering
	\label{tab:recon_table}
    \begin{tabular}{lrrr}    
        \toprule
        &         MH-Wiener &  S-Wiener &     Z-Wiener \\
        \midrule
        VEM       &  \bf{11.8} &    11.4 &  11.4 \\
        Heuristic &   \bf{6.4} &     5.8 &   5.8 \\
        MCEM      &  11.9 &          / &        / \\
        \bottomrule
    \end{tabular}
    \caption{SDR (dB) as a function of the reconstruction method. MH-Wiener, S-Wiener and Z-Wiener methods are defined in Section \ref{ssec: Speech recon}}
\label{table:recon_table}
\end{table}
\vspace{-1cm}

\section{Conclusion}
In this paper, we proposed a speech enhancement method based on a variational EM algorithm. Our main contribution is the analytical derivation of the variational steps in which the encoder of the pre-learned VAE can be used to estimate the variational approximation of the true posterior distribution, using the very same assumption made to train VAEs. Experimental results showed that the principled approach outperforms its heuristic counterpart, and produces results on par with the algorithm proposed by Leglaive et al. \cite{Leglaive2018Single} in which the true posterior is approximated using MH sampling. Additionally, the proposed algorithm converges 36 times faster. Future work includes multichannel and multisource extension of this work. The use of other statistical models of speech spectra \cite{nugraha2019deep, Magron2018ani, Liutkus2018donut} will also be explored. 

\bibliographystyle{IEEEtran}

\bibliography{mybib}

\end{document}